\documentclass[preprint,12pt]{elsarticle}

\usepackage{amssymb}
\journal{the arXiv.org}

\begin{document}

\begin{frontmatter}

\title{Intellectual  Management of Enterprise }

\author{Yuriy Ostapov}
\address{Institute of Cybernetics of NAS of Ukraine, 40 Acad. Glushkova  avenue,  Kiev, Ukraine.
E-mail: yugo.ost@gmail.com }

\begin{abstract}

A new technology (in addition to ERP) is proposed to provide an increase of profit and  normal cash flow.
This technology involves the next functions: forming  of intellectual interface on a natural language  to communicate with  a control system;
joint planning of production and sales to get the maximal profit; an adaptation of control system to internal and external events.
The use of the natural language permits to overcome a barrier between the control system and upper managers. To solve posed actual problems of management
the selection of information from a database and call to mathematical methods are executed automatically.
Optimal planning provides the maximal use of available resources and opportunities of market.
Adaptive control implements the efficient reaction to critical events that  lead up to a decrease of profit and increase of accounts receivable.
 
\end{abstract}

\end{frontmatter}

\section{Introduction}

The given paper is devoted to actual problems connected with Enterprise Resource Planning (ERP) \cite{Chase, Olear}:

\begin{itemize}

\item   {\it the effective direct interaction of top managers with ERP-system;
\item  joint planning of sales and production;
\item  adaptive control of enterprise to provide reactions to internal and external  critical events}.
                                                                                  
\end{itemize}

Considering a question of ERP-system recoupment, it can assert that such system  is  {\it necessary but not sufficient}   to provide effective business \cite{Gold3}.   
Thus, we propose a new technology {\it Intellectual  Management of Enterprise}  (in addition to ERP)  that realizes the next functions:

\begin{itemize}

\item   {\it the use of natural language} to communicate  with  control system\footnote {Hereafter we shall use the term "control system"  to mean "computer control system  of enterprise".} ;
\item   {\it the prediction and analysis of sales} on the base of modern mathematical methods;
\item   {\it optimal planning} to get the maximal profit considering constraints of resources and the prediction of sales;
\item   {\it the adaptation of control system}  to critical events that exert a significant effect on sales and production;
\item   {\it the use of natural language understanding algorithms}  to solve management problems connected with the application of text information about organizations, persons, and goods.                                                                                  
\end{itemize}

A model of enterprise presented by means of a  control system database  can serve only a starting point to take  decisions.
Algorithms of control must be founded on  inference rules, which generalize results of researches in enterprise management theory \cite{Chase, Drury, Horn, Kotler, Malhot}. 

When top managers use a natural language to communicate with control system, a process of decision making goes to the fundamentally other level as a barrier between the control system and top managers disappears.
It is of first importance as  such managers use economical categories to control an enterprise: effectiveness, profitability, stability \cite{Drucker}.  
When answers for managers are formed, algorithms of mathematical analysis are called {\it  automatically}, and results are displayed  in a convenient form.

According to {\it  theory of constraints}\cite{Gold1}, planning of production must consider "weak places".
A proposed approach is based on optimal planning taking into consideration constraints of resources and the prediction of sales.

An enterprise management must  react to critical events that break a normal work of production and realization of goods.
It is necessary routinely to estimate a state of plan targets and to correct these plans.
Furthermore, constant analysis of critical events is required to assess consequences of such events and to propose appropriate decisions.

\section{Model of market}

 A model of market includes descriptions of organizations (competitors,  buyers,  wholesalers,   retailers) --- financial state, marketing and production problems. 
In addition to organizations, it is necessary to have descriptions of  goods\footnote {Only for goods that belong to the same group as for considered enterprise.}:

\begin{itemize}

\item  sales  --- volume, price, producer, buyer, time of sale;
\item characteristics;
\item  documents containing information about goods.
                                                                              
\end{itemize}

Except  formalized information about organizations and goods, text information from documents, mass media, and Internet can be as well of great interest. 
Such information can be used for management if semantics of these texts will be adequately presented in a control system. 
Such representation of semantics is realized with the help of natural language understanding algorithms \cite {Ostapov1, Ostapov2, Ostapov3}.

 When we model a market, two sorts of market should be distinguished:

\begin{itemize}

\item   {\it consumer market}  ---  sales of goods for people;
\item   {\it  industrial market}  ---  sales of goods for enterprises.
                                                                              
\end{itemize}

\subsection{Consumer market }

 A price of consumer goods is determined by a retailer.
Analysis  of {\it  consumer value, segmentation} and {\it  positioning} is executed with methods of mathematical statistics.
Goods  are described by means of {\it  characteristics}, which are measured with appropriate {\it  scales} (nominal, ordinal, interval and  ratio)\cite {Malhot}.

 {\it  A consumer value} is determined with the expression:
\begin{equation}
   Q = \sum_{i=1} ^ {m} w_{i} \,  x_{i}  ,                                                            
\end{equation}
where   $x_{i}$   is  a characteristic of goods exerting influence on the consumer value. Parameters of model  $w_{i}$  are calculated with the help of  {\it  the least-squares method} 
using results of inquiries and consumer opinions\cite {Malhot}.

To study a demand  a mathematical model  is built with the help of data processing methods. For example, it can apply the linear equation of regression
 connecting the change of sale volume $ \Delta V$ with  changes of goods characteristics  $\Delta  x_{i}$ :
\begin{equation}
   \Delta V= \sum_{i=1} ^ {m} c_{i} \,  \Delta  x_{i} ,                                                               
\end{equation}
The most essential properties of goods, which   have the correlation with sale volume, are used as characteristics $x_{i}$
(for example, price, consumer value, purchasing power). Parameters of model  $c_{i}$  are calculated with the least-squares method 
using data about sales\cite {Malhot}.

The equation (2)  is valid for a small interval of prices. In the general case, a nonlinear model is more adequate:
\begin{equation}
   V= f (x_{1},...,x_{n})  .                                                                
\end{equation}

The functional dependence $f$ can  be continuous or have discontinuity. Such general dependences are realized by means of algorithms of nonlinear regression analysis
and other algorithms of data processing\cite {Kantar, Malhot}. For example, it can use methods of pattern recognition with the help of separation into classes for an interval of V.
Learning of model is implemented on the base of sale information.

High reliability of prediction is provided, first, with the sufficient number of data, second, with the correlation between V and $x_{i}$, third,
with the small dispersion of V for given $x_{i}$\cite {Malhot}.

\subsection{ Industrial market}

If the number of buyers-organizations is little,  {\it an algorithm of purchase} can  be use  to analyse and manage sales.
Let us consider a purchase  of goods  X  by an organization A. These goods are produced by an organization B.
An algorithm of such purchase is:\\[10pt]
\hspace*{20pt} 1. If the goods X answer demands of the organization  A, we go to step  2, otherwise to step 3.\\
\hspace*{20pt} 2. If  the organization A has  a purchasing arrangement for analogous goods, we go to step  4, otherwise to step 5.\\ 
\hspace*{20pt} 3.  If the organization  B can adapt the goods X to demands of the organization  A,  we go to step  2, otherwise to step 9.\\
\hspace*{20pt} 4.  If the organization  A has problems with delivery  and quality for the analogous goods, we go to step  5, otherwise to step 9.\\
\hspace*{20pt} 5.  If the organization  A is a reliable client\footnote {We are dealing with responsibility of A.}, we go to step  6, otherwise to step 9.\\ 
\hspace*{20pt} 6.  If the organization  A agrees to proposed price, we go to step  8, otherwise to step 7.\\
\hspace*{20pt} 7.  If the organization  B  brings down the price, we go to step  6, otherwise to step 9.\\
\hspace*{20pt} 8.  A bargain can be concluded.\\ 
\hspace*{20pt} 9.  A bargain is not concluded.\\[10pt]

\section{Analysis of delivery chain for consumer goods}

Analysis of delivery chain plays a leading role in enterprise management \cite {Chase}.
Consider the simple delivery chain that involves {\it  a producer, wholesalers, retailers}, and {\it  buyers}.
For such delivery chain there are three prices for certain goods: {\it  basic, wholesale}, and {\it  retail}.
We suppose that the producer sets one basic price for all the wholesalers.
This price is established in terms of the next factors:

\begin{itemize}

\item ensuring profit;
\item availability of competitors;
\item a stage of life cycle  for goods;
\item a consumer value.
 
\end{itemize}

A wholesaler quotes one price for all the retailers. This price is founded on the basic price and 
takes account of the customer demand and expenses. 

For  members of chain it can formulate the next factors that exert influence on their activity:

\begin{itemize}

\item size of stocks  and their dynamics;
\item prices of competitive goods;
\item rating of goods;
\item customer complaints.
 
\end{itemize}

\subsection{Prediction of retailer actions}

It can describe the next rules for retailer actions in a considered period:\\[10pt] 
\hspace*{20pt} 1.  Let us find a dependence between a price and volume of sales for certain goods (as indicated above).\\
\hspace*{20pt} 2.  Let us set the price and supposed volume of sales for this goods taking into account expenses and planed profit.\\ 
\hspace*{20pt} 3.  If stocks of these goods are not sufficient for sales, then we determine size of order for a wholesaler.  \\[10pt] 

The retailer profit (for the realization of certain goods party) is determined with the formula:
\begin{equation}
   \Pi =  P_{s} \,V_{s} -  P_{d} \,V_{s} - I_{s} ,                                                           
\end{equation}
where   $P_{s}$   is a retail price, $V_{s}$   is a volume of sales, $P_{d}$   is a wholesale price, 
$I_{s}$ is  expenses of retailer connected with the given goods party. 

\subsection{Prediction of wholesaler actions}

It can point the next rules for wholesaler actions in a considered period:\\[10pt] 
\hspace*{20pt} 1.  If there is a long-term agreement about  delivery to a retailer, and stocks of goods are sufficient, 
then an order is realized  according to this agreement. If  stocks of these goods are not sufficient, we go to step  3.  If there is not the long-term agreement, we go to step  2.  \\
\hspace*{20pt} 2.  If stocks of goods are sufficient for the realization of retail orders, then  delivery from the producer is not demanded.  If stocks of goods are not sufficient, we go to step  3. \\ 
\hspace*{20pt} 3.   Let us determine order size for the producer in terms of stocks and available (or supposed) orders from retailers. \\[10pt] 

The wholesaler profit (for the realization of certain goods party) is determined with the formula:
\begin{equation}
   \Pi =  P_{d} \,V_{d} -  P_{e} \,V_{d} - I_{d} ,                                                           
\end{equation}
where  $P_{d}$   is a wholesale price, $V_{d}$   is a volume of sales,  $P_{e}$ is a basic price,
$I_{d}$ is expenses of wholesaler connected with the given goods party. 

\subsection{Prediction of producer actions}

It can describe the next rules for producer actions:\\[10pt] 
\hspace*{20pt} 1.  If there is a long-term agreement about  delivery to a wholesaler in a considered period, and stocks of goods are sufficient, 
then an order is realized  according to this agreement. If  stocks of these goods are not sufficient, we go to step  3.  If there is not the long-term agreement, we go to step  2.  \\
\hspace*{20pt} 2.  If stocks of goods are sufficient for the realization of wholesale orders, then the production of goods  is not demanded.  If stocks of goods are not sufficient, we go to step  3. \\ 
\hspace*{20pt} 3.  Let us determine a production volume for the given goods in terms of stocks and available (or supposed) orders from wholesalers. \\[10pt] 

\section{ Analysis of delivery chain for industrial goods}

For industrial goods the direct connection between a producer and a buyer  is used sufficiently often. If there are a lot of buyers, then it can use statistical methods.
If there are few buyers, then it can apply {\it an algorithm of purchase} (see above).

If there are wholesalers,  two cases should be distinguished. If a wholesaler has a lot of buyers,  it can use statistical methods.
A wholesale and basic price for certain goods are determined with the same factors as for consumer market. If a wholesaler has few buyers,  it can apply the algorithm of purchase.

\section{Analysis and management  of sales}

Analysis of sales is demanded when it is necessary to find causes of reduction for sales. Basic causes are:

\begin{itemize}

\item  service problems;
\item worsening of relations with wholesalers and retailers;
\item availability of obsolete and poor goods;
\item an appearance of competitive goods;
\item a decrease of purchasing power.
\end{itemize}

An answer to a question about  causes of sale reduction is based on analysis of these factors.

To find an answer to the question  {\it What characteristics are the most essential for pointed goods?} one must establish the connect between basic characteristics of these goods
and a volume of sales using correlation and variance analysis\cite {Malhot}.

To form an answer to the question  {\it What customer complaints are there for pointed goods?} one must  find  all  opinions, complaints, censorious remarks of customers and establish if action is taken
to exclude defects.

To get an answer to the question  {\it What service problems  are there  for pointed goods?} it is necessary to  explore  all censorious remarks about service and  to check if action is taken.

Consider as well problems of sale management. Management  always is begun  with planning.
On this stage we must maximally take account of market opportunities.
It should be distinguished industrial and consumer market.
In the first case, if a planed volume of sales is not provided with  available orders, a question arises:
{\it What else buyers can   conclude a bargain?}. To form an answer all supposed buyers are checked, and possibility 
of such bargain is estimated using the above-mentioned algorithm of purchase.

In the case of consumer market, to prepare an answer to the question {\it How to increase sales?}
one must form propositions for:
\begin{itemize}

\item improvement of  service and quality for goods ;
\item a change of  volumes and prices (taking into account competitive prices and opportunities of production);
\item a change of sale organization.
 
\end{itemize}
 
To form an answer to the question about effectiveness of marketing it can use methods of correlation and variance analysis\cite {Malhot}.

\section{Optimal planning} 

Planning in living control  systems  is based on analysis of demand or available orders.
Consider the first case when planning is founded on consumer demand, and volumes and prices can be changed  in certain limits.
In this case, planning of sales must provide the maximal income. Planning of production is implemented in terms of sale plan.

The next approach is more effective. Taking into account the prediction of sales, normative expenses, and resource constraints,  we find {\it such allocation of production volumes  and sale prices
that provides the maximal profit}. 

Let us go to exact definitions. An expression for operation profit is:
\begin{equation}
   \Pi =  \sum_{i=1} ^ {n} p_{i} \,  v_{i}  -   \sum_{i=1} ^ {n} a_{i} \,  v_{i}  -   a_{0}, \, y_{i} = v_{i}-z_{i}, i=1,...,n,                                                  
\end{equation}
where $n$ is the number of goods types, $p_{i}$   is a price for goods with the index  $i$, $v_{i}$  is a volume of sales for goods with the index  $i$,
$y_{i}$   is a volume of production for product (goods)  with the index  $i$, $z_{i}$   is stocks for product with the index  $i$, 
$a_{i}$   is direct expenses for product with the index  $i$,  $a_{0}$   is  indirect expenses connected with the production of considered product parties.
 
Meaning of equation (6) is that  a total product cost is subtracted from  a general income to find a profit\cite {Drury}. 
The price and volume of sales can be connected with the equation (2) or (3). In general case, the model of delivery chain can be used (see above).

The problem of optimal planning consists in the determination of sets $p_{1},...,p_{n}$ and $v_{1},...,v_{n} $ providing the maximal profit $\Pi$ taking account of constraints
for market and production. The market constraints are determined with inequations:
\begin{equation}
    p_{imin} \,\leq p_{i} \,\leq  p_{imax}, \,  v_{imin} \,\leq v_{i} \,\leq  v_{imax}, \,i=1,...,n .                                                 
\end{equation}

To consider constraints for joint used resources first we find the duration of maximum long operation for making of product  with index  $k$:

\begin{equation}
     \tau_{k} = \max \sum_{j=1} ^ {r}t^{k}_{ij},  \, i=1,...,m,                                                 
\end{equation}
where $t^{k}_{ij}$ is a duration of an operation with index $i$ for  making of a component with index $j$ if this component is used for the product with index  $k$.
If the expression (8) is executed for $i=i_{k}$, then such operation with the index $i_{k}$ is named  "a weak place" (or "a weak link")
 \footnote { Such approach was first proposed by E.Goldratt \cite {Gold1}. For simplicity, we suppose that the considered operation is executed  only at one workplace.}.

For products with indexes $j_{1},...,j_{r}$ that have "general weak place" $i_{0}$, the next condition must be executed:
\begin{equation}
     \tau_{j_{1}}\,y_{j_{1}} + ... + \tau_{j_{r}}\,y_{j_{r}} \, \leq \,T  ,                                          
\end{equation}
where $T$ is the general work time for planed period. We suppose that the operation $i_{0}$ is not used for making of other products.

To solve the task of optimal planning it can use the method of statistical testing. 
Optimal planning will be successful at the next condition:\\[10pt]
\hspace*{20pt} 1. There are  carefully checked norms of direct expenses for all  production operations as well as norms of use  for materials and energy supply. \\  
\hspace*{20pt} 2. The equation of demand describes reasonably well the dependence between a volume and price of sales. \\ 
\hspace*{20pt} 3. Indirect expenses are distributed to functions (method ABC)\cite {Drury}. \\[10pt] 

\section{Analysis of financial results}

In spite of availability of ERP-system, which provides planning and subsequent control, top executives feel a  need directly to communicate with control system.
Such communication on natural language permits to get answers to actual questions about economical state of own enterprise and   causes of  this state worsening.
It is essentially that such answers are formed automatically. This allows  to check actions of subordinate structures.

To find causes of financial state worsening for own enterprise such questions can be used:\\[10pt] 
{\it \hspace*{20pt} What causes  are there for decrease of profit   in a given period?\\
\hspace*{20pt}  Why is a product cost increased in a given period?\\
\hspace*{20pt} Why are sales decreased in a given period?}\\
 
If   an organization has  branches,  top executives can  take an interest in economical state of these companies. 
If  a database of control system contains the next data for branches :
\begin{itemize}

\item about expenses for certain sorts of production;
\item  about sales of certain goods (including problems of quality and service),
 
\end{itemize}
then it can get answers to above-mentioned  questions about causes  of financial state worsening.

Furthermore, it is necessary to get operative information about  financial state of competitors, partners, branches (on the base of standard financial  statements).
Such information plays  a great role for decision making.

Answers to questions about a certain organization:\\[10pt] 
{\it \hspace*{20pt} What  are  general financial results in a given period?\\
\hspace*{20pt} What is profit  in a given period?\\
\hspace*{20pt} What  is cash flow in a given period?\\
\hspace*{20pt} What  is general accounts receivable in a given period?}\\[10pt] 
permit  upper managers  to take appropriate action. To provide these functions one must save  standard financial  statements for considered organizations in the database of control system.

\section{Adaptive control}

As noted above, a control system of enterprise must react to  external and internal critical events. 
First of all, one must execute the regular estimation of production state  to explore breaches of plan and then to correct plan targets taking into account resource constraints.
Furthermore,  algorithms of adaptation include the analysis of information about events that break normal industrial process and realization of goods.
Descriptions of such events are inputted into control system on a natural language.

If an event applies to production process, then the estimation of consequences is executed, and propositions for decision making are  formed.
 If there are serious problems with delivery of components, materials, and  energy supply, then it is necessary to take appropriate action and to correct plans (as the need arose).
When competitive goods  or new opportunities for sales appear, one must  correct  a plan of sales.

For large-scale enterprises, which usually  realize production in many regions, it is necessary to  react to political events, ecological disasters, changes of macroeconomic factors in these regions
if such events can  break a living delivery chain and decrease sales.

General meaning of adaptive control implies that one must  provide  efficient reaction to events and   take  basic management decisions  only in terms of all available  information about critical events for market and production.
It should be emphasized that the greater is a scale of production, the greater meaning has  message handling  service  for internal and external events. 

To describe algorithms of decision making connected with critical events an inference engine  is applied in a control system.
This engine realizes a direct and reverse inference\cite {Ostapov4}. 

The direct inference is executed with the help of decision tree. The reverse inference uses hierarchy of purposes, which is built on the base of inference rules.
Predicates of rules can be complex and simple. Simple predicates implement arithmetic and logical operations, the selection from  a database as well as operations of pattern recognition and modeling \cite {Ostapov4}. 

\section{Experimental system}

To check above-mentioned control algorithms  an experimental system was created. MS SQL Server 7.0 is used as a database management system. 
A language for communication is Russian. A database describes an enterprise activity. This enterprise assembles notebooks. 
The database includes such tables:
\begin{itemize}

\item {\it balans} --- for balance;
\item  {\it money} --- movement of funds;
\item  {\it finres} --- financial results;
\item  {\it logist} --- the description of logistics and sales;
\item  {\it docum} --- the description of documents connected with goods realization;
\item  {\it tovar} --- the description of goods;
\item  {\it character} --- the description of goods characteristics;
\item  {\it calcul} --- the description of production operations;
\item  {\it debitor} --- the description of accounts receivable.
 
\end{itemize}

Furthermore, there are as well dictionaries to realize algorithms of natural language understanding\cite {Ostapov1}.
The experimental system implements such functions:

\begin{itemize}

\item analysis of financial state;
\item forming of propositions for improvement of economical state;
\item optimal planning;
\item analysis of sales.

\end{itemize}

\subsection{Analysis of financial state}

This function permits to find answers to actual questions about economical state of the enterprise and causes of  this state worsening.
For example, for the question {\it What causes are there for increase of product cost in second quarter, 2010?} \footnote
{Hereafter all questions and answers are given in translation from Russian.}, we have such answer:

{\it The prime cost of the notebook TN20A is more than the sale price as:\\[10pt] 
\hspace*{20pt} a mother board has a retail price more than the average  market price;\\
\hspace*{20pt} a hard disk has a retail price more than the average  market price;\\
\hspace*{20pt} a screen matrix has a retail price more than the average  market price;\\
\hspace*{20pt} the operation "assembly of notebook" is executed with productivity less than norm.}\\[10pt]

\subsection{Forming of propositions for improvement of economical state}

This function provides forming of propositions for  manager on the base of economical analysis.
For example, for the question {\it How to  increase the income from realization of production in third quarter, 2010?}, we have the answer:

{\it Results of economical analysis show that:\\[10pt] 
\hspace*{20pt} the notebook TN20A is in need of quality improvement;\\
\hspace*{20pt} negative opinions are given in the sheet of documents;\\
\hspace*{20pt} the notebook TN20A keeps less than  50 percents of market, therefore it can increase  a realization volume at the expense of other regions;\\
\hspace*{20pt} the notebook TN301 keeps less than 50 percents of market, therefore it can increase  a realization volume  at the expense of  other regions.}\\[10pt]

\subsection{Optimal planning}

This function provides optimal planning as considered above. To form optimal volumes of production the next operations are executed:
\begin{itemize}

\item for produced goods --- a dependence between a volume and price of sales is found on the base of information about sales (in the database);
\item for each variant of plan --- direct and indirect expenses are determined for all products, and then a profit from sales is calculated;
\item optimal variant corresponds to the maximal profit.

\end{itemize}

\subsection{Analysis of sales}

Analysis of sales is executed as indicated above. For the question {\it How to increase  the volume of sales  in third quarter, 2010?}, 
an  answer is:\\

{\it Results of marketing analysis show next:\\[10pt] 
\hspace*{20pt} the notebook TN20A must be modernized as:\\
\hspace*{20pt} size of RAM is not sufficient;\\
\hspace*{20pt} the notebook TN20A keeps less than  50 percents of market, therefore it is necessary to increase sales  at the cost of  other regions;\\
\hspace*{20pt} the notebook TN301 keeps less than 50 percents of market, therefore it is necessary to increase  sales  at the cost of  other regions}\\[10pt]

For the question {\it What characteristics  are best suited to notebooks of  budget class in  2010?}, 
an  answer is:\\

{\it \hspace*{1pt} size of RAM  is 2 Gb;\\
\hspace*{20pt}  a price is 4048 gr;\\
\hspace*{20pt} an operation system is DOS.}\\[10pt]

\section{Conclusion}

Using modern progress in artificial intelligence, it can  realize new functions for enterprise management (in addition to ERP):
\begin{itemize}

\item   forming  of intellectual interface on a natural language to communicate with  a control system;
\item   joint planning of production and sales to provide the maximal profit;
\item   the adaptation of the control system to internal and external events.

\end{itemize}

Proposed new functions  are the most actual for medium and large-scale enterprises.
The next factors have more important meaning for the use of these functions:
\begin{itemize}

\item   availability of  branches in other regions; 
\item   a large range of products;
\item   spirited competition;
\item   a delivery chain involving a lot of regions.

\end{itemize}

Proposed technology serves in order to improve  management effectiveness, to increase profit, and to provide  normal cash flow.
It is implemented with the help of the next factors of successful  business:

\begin{itemize}

\item   the close interaction of upper managers with the control system that permits in time to message about basic causes of profit decrease (to take action); 
\item   optimal planning of production oriented on maximal profit (considering  production constraints and using the valid prediction of sales);
\item   ensuring efficient reaction to critical events that lead up to decrease of profit and increase of accounts receivable.

\end{itemize}

\end{document}